\documentclass[useamsfonts,dvipdfmx]{pasj00}

\begin{document}
\SetRunningHead{T. Okamoto et al.}{Cosmic evolution of simulated bars}

\title{Cosmic evolution of bars in simulations of galaxy formation} 
\author{Takashi \textsc{Okamoto}$^1$ %
}
\affil{$^1$Department of Cosmosciences, Graduates School of Science, Hokkaido University, N10 W8, Kitaku, Sapporo, 060-0810, Japan}
\email{okamoto@astro1.sci.hokudai.ac.jp}

\author{Mari \textsc{Isoe}$^{2, 3}$}
\affil{$^2$Department of Astronomy, School of Science, The University of Tokyo, 7-3-1 Hongo, Bunkyo-ku, Tokyo 113-0033}
\affil{$^3$Division of Theoretical Astronomy, National Astronomical Observatory 
of Japan, 2-21-1 Osawa, Mitaka, Tokyo 18l-8588}
\email{isoe.mari@nao.ac.jp}
\and
\author{Asao \textsc{Habe}$^{1}$}
\email{habe@astro1.sci.hokudai.ac.jp}

%


\maketitle

\begin{abstract}
We investigate the evolution of two bars formed in fully 
self-consistent hydrodynamic simulations of the formation of 
Milky Way-mass galaxies.  
One galaxy shows higher central mass concentration and has a 
longer and stronger bar than the other at $z = 0$. 
The stronger bar evolves by transferring its angular momentum 
mainly to the dark halo. Consequently the rotation speed of the 
bar decreases with time, while the amplitude of the bar increases 
with time. 
These features qualitatively agree with the results obtained 
by idealized simulations. 
The pattern speed of the stronger bar largely goes up and down 
within a half revolution in its early evolutionary stage. 
These oscillations occur when the bar is misaligned with 
the $m = 4$ mode Fourier component. 
These oscillations correlate with the oscillations in the 
triaxilality of the dark matter halo, but differently from the 
way identified by idealized simulations. 
The amplitude of the weaker bar does not increase despite the fact 
that its rotation slows down with time.
This result contradicts what is expected from idealized 
simulations and is caused by the decline of the central density 
associated with the mass loss and feedback from the stellar 
populations. 
The amplitude of the weaker bar is further weakens by the angular 
momentum injection by the interactions with stellar clumps in the 
disk. In the both galaxies, the bars are terminated around the 
4:1 resonance.  
\end{abstract}

\section{Introduction}

Roughly two-thirds of disk galaxies have bars, of which half have strong 
bars today \citep{eskridge00, barazza08}.  
The bar fraction seems to be 
a strong function of redshift; it decreases by a factor of three 
from $z = 0$ to $z = 0.8$ \citep{sheth08}.  
Bars can affect galaxy properties  
by driving secular evolution 
(e.g. \cite{kormendy79, combes81, pfenniger90, combes93, 
debattista04, kk04, athanassoula05, athanassoula13, kormendy13}), 
and thus a number of simulations have 
performed to study their formation and evolution processes. 

Most of the simulations employ idealized initial conditions for isolated 
galaxies in order to answer specific questions (e.g.
\cite{combes90, debattista00, athanassoula03, martinez-valpuesta06, amr13}).  
A general picture obtained by these studies is that a bar continuously grows 
and slows down with time as its angular momentum gets transferred to the 
dark halo and the outer disk
(e.g. \cite{weinberg85, combes93, athanassoula03}).

In reality however disks and dark halos keep growing while bars 
form and evolve. 
The dark halos resulting from cosmological simulations  are 
naturally triaxial \citep{frenk88, jing02} and abundant in substructure 
\citep{moo99, kly99, aquarius}, 
both of which should exert torque on bars. 
Galaxy interactions could trigger bar formation \citep{miwa98, berentzen04}.
Moreover, even disk orientation changes with time owing to the misalignment 
of the angular momentum vector of the newly accreting gas 
\citep{oka05, okamoto13}.

Early attempts to investigate the effects of cosmological growth of galaxies 
on the bar evolution are made either by simplifying the initial density 
perturbations \citep{heller07, romano-diaz08} or by embedding a live disk 
in a growing dark halo \citep{curir06}.  
\citet{kraljic12} use a cosmological $N$-body simulation as boundary 
conditions for sticky particle simulations and follow galaxy and bar 
evolution in evolving halos. 
They confirm that the bar fraction in fact decreases with increasing 
redshift. 
\citet{scannapieco12} analyze the bars in fully self-consistent 
simulations of galaxy formation at $z = 0$ and compare their properties 
with those in the idealized simulations. 

Only recently have cosmological simulations with sufficient resolution to 
follow evolution of detailed structure in disk galaxies become possible 
\citep{eris, okamoto13, magicc, marinacci14}. 
An important difference of these simulations from the idealized simulations is 
that they invoke much stronger stellar feedback than the idealized simulations 
normally assume, otherwise simulations form 
too many stars for a given halo mass (e.g. \cite{aquila, okamoto13, okamoto14}). 
The feedback is so strong that it can lower the central stellar and dark matter 
density during the galaxy evolution \citep{duffy10, governato12}, and hence 
orbits of stars should be affected. 

Two Milky Way-mass galaxies formed in the cosmological simulations by 
\citet{okamoto13} offer a unique opportunity to investigate the evolution 
of bars in the context of the $\Lambda$-cold dark matter ($\Lambda$CDM) 
cosmology. 
One galaxy is clearly barred from $z \simeq 1$ to $0$, while the other has a 
much weaker bar.  
In this paper we compare and contrast these two galaxies focusing on the 
evolution of their bars. 
We also compare their properties with those obtained by idealized simulations.  

This paper is organized as follows. In section~2 we briefly describe the 
simulations and galaxies we analyze. We present our results in section~3 
We then  discuss them and summarize our main conclusion in section~4. 


\section{Sample galaxies}

We study the evolution of the bars formed in two smoothed particle hydrodynamic 
(SPH) simulations of galaxy formation in a $\Lambda$CDM universe.  
The cosmological parameters employed in these simulations are: 
$\Omega_0 = 0.25$, $\Omega_\Lambda = 0.75$, $\Omega_\mathrm{b} = 0.045$, 
$\sigma_8 = 0.9$, $n_\mathrm{s} = 0.9$, and a Hubble constant of 
$H_0 = 100~h$~km~s$^{-1}$~Mpc$^{-1}$, where $h = 0.73$. 
The galaxies are selected from a cosmological periodic box of a side 
length of $100~h^{-1}$~Mpc and called `Aq-C' and `Aq-D' according to 
the labeling system of the Aquarius project \citep{aquarius}.  

The dark matter particle masses are $2.6 \times 10^5$ and 
$2.2 \times 10^5$~M$_\odot$ for Aq-C and Aq-D, respectively, and 
the original SPH particle masses are $5.8 \times 10^4$ and 
$4.8 \times 10^4$~M$_\odot$, respectively. 
At $z < 3$ the gravitational softening lengths are fixed in physical 
coordinates as $\epsilon = 0.257$ and $0.240$~kpc respectively in 
Aq-C and Aq-D. The simulations include radiative cooling, photo-heating 
by the ultra-violet background \citep{ogt08}, star formation, timed release 
of energy, mass, and metals by type II and Ia supernovae and AGB stars 
\citep{ofjt10, okamoto13}. 
It should be noted that the mass resolution of our simulations are as 
good as that in the high resolution idealized simulations, but the 
spatial resolution is still poor compared with that of the recent 
idealized simulations ($50$~pc in \cite{amr13}). 

The way of implementing feedback is a key to reproduce observed properties 
of galaxies \citep{ofjt10, okamoto14}. 
In our simulations, the energetic feedback from supernovae is modelled
as winds. A gas particle may receive an amount of energy, $\Delta E$, form 
type II supernovae during a time-step, $\Delta t$. 
We add this gas particle to winds with a probability, 
$p_\mathrm{w} = \Delta E/[(1/2) m_\mathrm{gas} v_\mathrm{w}^2]$, 
where $m_\mathrm{gas}$ is the mass 
of the gas particle and $v_\mathrm{w}$ is the initial wind speed. 
The initial wind speed, $v_\mathrm{w}$ is given as $v_\mathrm{w} = 5 \sigma$, 
where $\sigma$ is the one-dimensional velocity dispersion of the dark matter 
particles around the gas particle. 

Doing this ensures that the wind mass generated by a type~II supernova
is proportional to $\sigma^{-2}$, and therefore less mass galaxies 
blow more winds per unit star formation than there more massive 
counterparts. 
This feedback model explains luminosities and metallicities of galaxies 
ranging form the Local Group satellites \citep{ofjt10} to more massive 
galaxies  \citep{okamoto14}. 
The supernova feedback implemented this way is the strongest among the 
simulations presented in \citet{aquila}, and the resulting stellar mass of 
the galaxies is consistent with what expected by the abundance matching for 
their halo masses.  \citep{guo10, behroozi13, moster13}. 

The simulations follow the evolution of the stellar populations throughout 
their entire life. Therefore the star particles lose their masses not 
only by the type II supernovae and mass loss from the massive stars but also 
by type Ia supernovae and the stellar mass loss from the intermediate mass 
stars. The feedback from old stellar populations by type Ia supernovae is 
also considered.

The stellar masses of these galaxies at $z = 0$ are 
$4.0 \times 10^{10}$~M${_\odot}$ (Aq-C) and 
  $3.1 \times 10^{10}$~M${_\odot}$ (Aq-D); 
hence their masses are close to that of the Milky Way. 
The global evolution and properties of these galaxies are fully 
described in \citet{okamoto13}. 
We here only show the evolution of the circular velocity profiles 
which is the most relevant to the current study. 

\begin{figure*}
 \begin{center}
  \includegraphics[width=\linewidth]{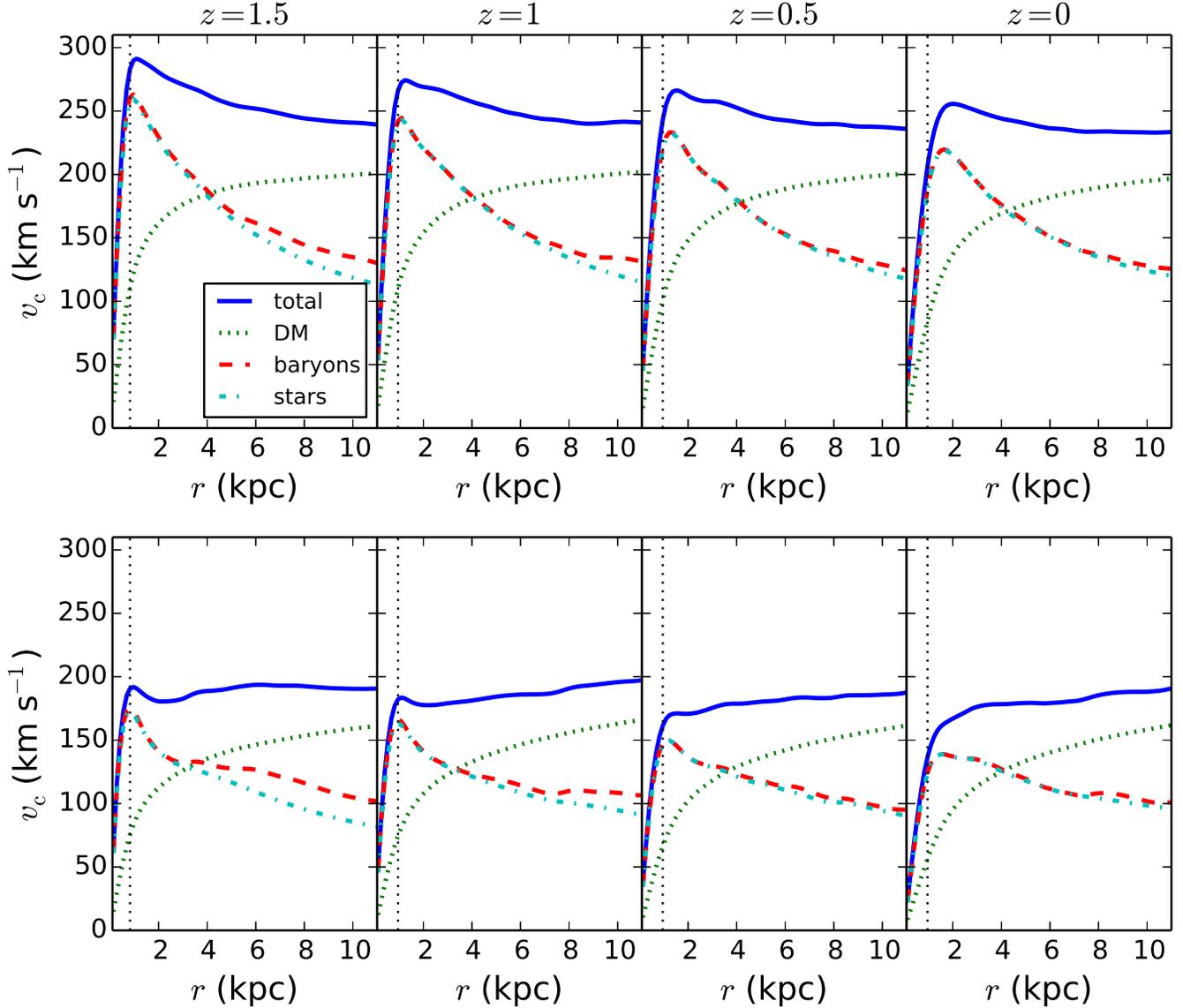} 
 \end{center}
\caption{
  Evolution of the circular velocity profiles. The upper and lower panels 
  show Aq-C and Aq-D, respectively.  
  From left to right, the circular velocities at $z = 1.5$, $1.0$, $0.5$, 
  and $0$ are plotted as functions of radius.  
  The blue solid lines are the circular velocities calculated from the 
  azimuthally averaged radial accelerations in the disk planes, and the 
  green dotted, red dashed, and cyan dot-dashed lines indicate 
  the contribution of the dark matter, baryons, and stellar components, 
  respectively. 
  The vertical dotted lines indicate the radii at which  
  $v_\mathrm{c}(r) = 1.1 v_\mathrm{c, b}(r)$, 
  where $v_\mathrm{c, b}(r)$ is the contribution of the baryonic matter to 
  the circular velocity at the radius, $r$.  
}
\label{fig:vc}
\end{figure*}
%
In order to characterize the mass distribution and the angular frequency of 
circular orbits, we calculate the circular velocity from the azimuthally 
averaged radial acceleration in the disk plane, $a_r(r)$,  
as $v_\mathrm{c}(r)^2 = - r a_r(r)$. 
We measure the radial acceleration at equally spaced 128 azimuthal positions 
at each radius in the disk plane to determine $a_r(r)$. 

We show the evolution of the circular velocity measured this way in 
figure~\ref{fig:vc}.  
We find that the central circular velocity continuously 
decreases with time in both galaxies. This is presumably due to the stellar 
mass loss and feedback since we do not find this behavior in simulations 
without stellar mass loss nor feedback.  
This drop changes the resonance structure as we will show later.  
We also find that Aq-C has more centrally concentrated mass distribution 
than Aq-D.   

The vertical line indicates the radius inside which a baryonic component 
dominates as $v_\mathrm{c}(r) < 1.1 v_\mathrm{c, b}(r)$, 
where $v_\mathrm{c, b}(<r)$ is the contribution of  
the baryons (gas and stars) to the circular velocity; 
the disks are bar unstable 
when this condition is satisfied \citep{eln93}, although this criterion 
is only appropriate for a pure $N$-body cold disk in a rigid halo 
(e.g. \cite{athanassoula08}). 
This radius is almost constant with redshift ($\simeq 1$~kpc) in both 
galaxies.  
Similarly, the radius at which the contribution of the baryons is equal 
to that of the dark matter is also constant with redshift ($\simeq 4$~kpc), 
indicating that the stellar mass fractions as functions of radius do not 
evolve strongly with redshift.  
Aq-C's circular velocity curve is more resembling to the universal 
rotation curve proposed by \citet{salucci07} than Aq-D's.  

\section{Results} 

In this section, we first describe how we identify stellar bars in the 
simulated galaxies and then undertake detailed studies of the evolution 
of the bars.  
In the following analyses, the $z$-direction is chosen to be parallel 
to the angular momentum vector of stars within 5 per cent of the virial 
radius\footnote{The virial radius is calculated based on the spherical 
collapse model \citep{ecf96}.} at given redshift unless otherwise stated. 
Note that the disk orientation significantly changes with redshift 
\citep{oka05, okamoto13}. 
We use stars with $|z| < 1~h^{-1}$~kpc to compute the surface stellar 
density to avoid the contamination of the stars in the satellite galaxies. 

\subsection{Bar identification}

The strength, length, and angle of a bar can be parameterized by the 
amplitude and phase of its Fourier component, defined by expressing the 
surface stellar density as a Fourier series, 
\begin{equation}
  \frac{\Sigma(r, \phi)}{\bar{\Sigma}(r)} 
  = 1 + \sum_{m = 1}^\infty A_m(r) \cos[m\{\phi - \phi_m(r)\}], 
\end{equation}
where $\Sigma(r, \phi)$ is the surface stellar density, $\phi$ is the azimuthal 
angle, $\bar{\Sigma}(r)$ is the azimuthally averaged surface stellar density at 
radius $r$, and $A_m$ and $\phi_m$ are the amplitude and phase of the 
$m$-th Fourier component, respectively. 
Before we perform this analysis, we remove stellar clumps in order for the 
$m = 2$ component not to include contributions of them, which are 
particularly abundant in Aq-D \citep{okamoto13}.   
The details of this procedure and its effect are described in 
appendix~\ref{app:clump}. In short, the removal of the clumps does not affect  
our results presented in this paper. 
We calculate the Fourier series for the face-on projections. 

Typically, the amplitude of the $m = 2$ mode, $A_2(r)$, has a peak if a 
bar exists. The phase, $\phi_2(r)$, should be constant in the bar 
region. 
We first identify the radius at which $A_2(r)$ takes the maximum value, 
$A_2^\mathrm{max}$, and we call this radius $r_2^\mathrm{max}$.  
The bar angle is defined as 
$\phi_\mathrm{bar} = \phi_2(r_2^\mathrm{max})$. 
We utilize the radial profile of the phase of the $m = 2$ Fourier 
component to determine the bar length. 
Outside the bar region, the phase should show large variation 
owing to a spiral structure or by the absence of clear structure.
We thus define the bar length, $r_\mathrm{bar}$, as the maximum radius 
where the bar angle 
$\phi_\mathrm{bar}$ and the phase of the $m = 2$ component differ by 
less than $\Delta \phi$. 
We employ $\Delta \phi = 0.1 \pi$ in this paper. 

\citet{am02} present a number of ways to measure bar length; 
each method has its own pros and cons. 
\citet{scannapieco12} compare three methods among them, which are readily 
applicable to disks obtained by cosmological simulations. 
We also compare these three methods in appendix~\ref{app:bar_length}; 
the first method is what we have described above, 
the second method uses the $m = 2$ amplitude profiles, and the 
third method compares the surface stellar density profile along the bar 
major axis with that along the bar minor axis. 
We find that the bar lengths obtained by these three methods agree 
reasonably well.

\begin{figure}
  \begin{center}
  \includegraphics[width=\linewidth]{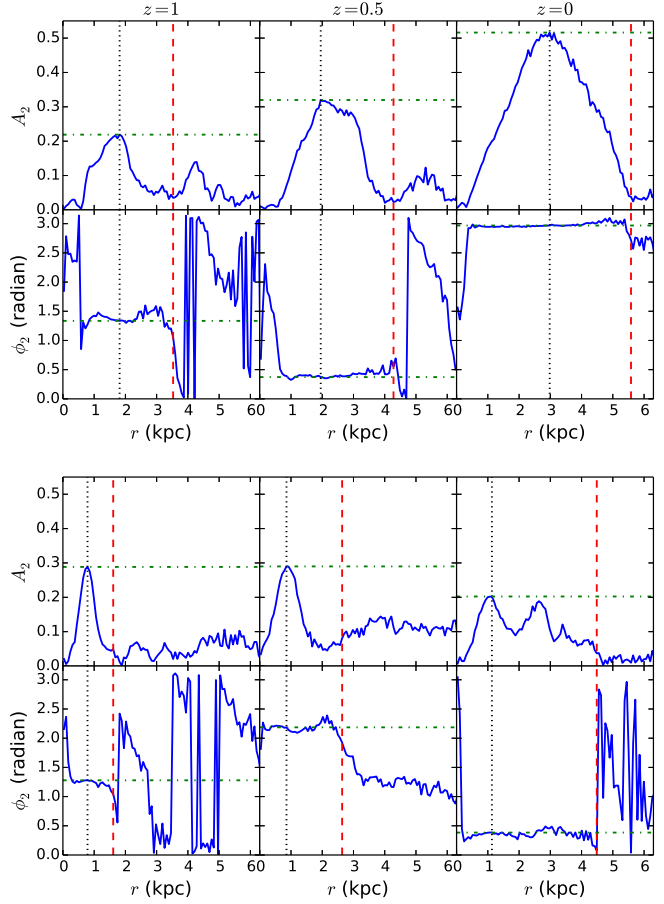} 
  \end{center}
\caption{The amplitude and phase profiles of the $m = 2$ mode. 
  The upper and lower six panels show Aq-C and Aq-D, respectively. 
  From left to right, the results at $z = 1$, 0.5, and 0 are presented.  
  The bar length, $r_\mathrm{bar}$, is indicated by the vertical red dashed 
  line in each panel. 
  The horizontal green dot-dashed lines indicate the values of 
  $A_2^\mathrm{max}$ and $\phi_\mathrm{bar}$ in the panels for amplitude and 
  phase, respectively.  
}
\label{fig:m2}
\end{figure}
%

We demonstrate our method in figure~\ref{fig:m2} where we plot the 
amplitude and phase profiles of the $m = 2$ Fourier component at $z = 1$, 0.5, 
and 0.  
We find that the amplitude, $A_2(r)$, has a clear peak 
at each redshift and the phase, $\phi_2(r)$, is nicely constant around the peak. 
We hence conclude that both $A_2^\mathrm{max}$ and $\phi_\mathrm{bar}$ are 
robustly defined by this method. 

On the other hand, the bar length, $r_\mathrm{bar}$, has weak dependence on 
the value of $\Delta \phi$, and thus it is less robustly defined\footnote{Any methods described in appendix~\ref{app:bar_length} introduce a tolerance 
parameter to define the bar length.}.  
In Aq-D, the amplitude, $A_2(r)$, has two peaks within the bar length 
at $z = 0$. 
Such a feature is hardly seen either in real galaxies (e.g. 
\cite{buta06, elmegreen07, buta09}) or in the idealized simulations 
(e.g. \cite{am02}). 
We therefore examine whether the outer peak is a different component from 
the inner component, which is coincidentally aligned with the inner component 
in appendix~\ref{app:outer}.

We find that the amplitude of the outer peak shows a rapid time variation
and the outer component always has almost the same phase as the inner 
component. 
By inspecting the surface density distribution, we speculate that 
the outer $m = 2$ peak is induced by interactions with the debris of 
the stellar clumps in the outer region of the bar, 
which is orbiting with the different angular velocity from the bar. 
We thus hold our definition of the bar length throughout this paper. 
We will later show how the bar length changes if we exclude the outer component. 
It should be noted that we always use the inner component to define the 
bar angle and amplitude of Aq-D, and therefore the treatment of 
the outer component does not affect most of the results and 
discussions presented in this paper. 

\begin{figure}
  \begin{center}
  \includegraphics[width=\linewidth]{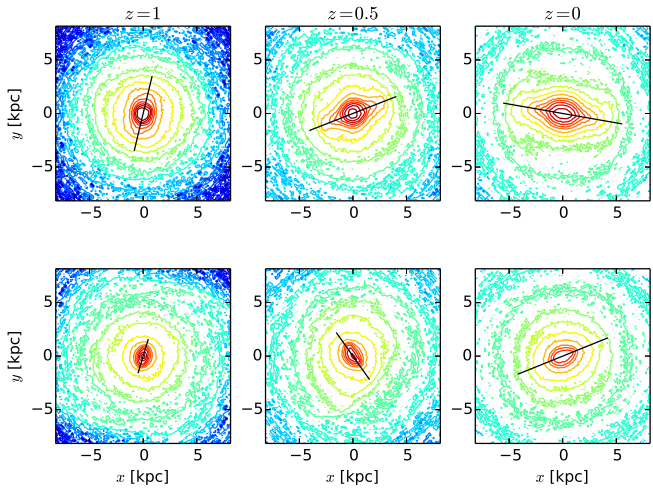} 
  \end{center}
\caption{
  Contour maps of the face-on stellar surface density of the galaxies. 
  We have removed the clumps from the density distributions as described 
  in appendix~\ref{app:clump}. 
  The upper and lower panels correspond to Aq-C and Aq-D, respectively, 
  and the galaxies at $z = 1$, 0.5, and 0 are shown from left to right. 
  The contour levels are logarithmic.  
  The black solid lines indicate the bars whose length and angle are defined 
  as described in the text. 
}
\label{fig:contour}
\end{figure}
%

The bars identified this way are overplotted on the face-on surface 
stellar density maps in figure~\ref{fig:contour}. 
The surface stellar density maps confirm that Aq-C always has a longer bar 
than Aq-D, while Aq-D has a better defined bar at $z = 1$, which is evident 
from figure~\ref{fig:m2} where the value of $A_2^\mathrm{max}$ of Aq-D 
at $z = 1$ is larger than that of Aq-C. 
We find that the central surface stellar density decreases with time 
in both galaxies. 
We also find that Aq-D has more structure than Aq-C such as spiral arms. 
Such structure in Aq-D is mainly induced by the interactions with the clumps 
(see appendix~\ref{app:clump}), 
which we have erased in figure~\ref{fig:contour}. 

The clumps sink to the central region \citep{okamoto13}, 
and it can change the bar properties by adding mass and changing 
the angular momentum of the central region. 
By comparing the bar lengths and angles indicated by the straight  
lines in figure~\ref{fig:contour} with 
the shapes of the isodensity contours, we conclude that our method 
identifies the bars in the simulated galaxies well.

\subsection{Redshift evolution}

\begin{figure}
  \begin{center}
  \includegraphics[width=\linewidth]{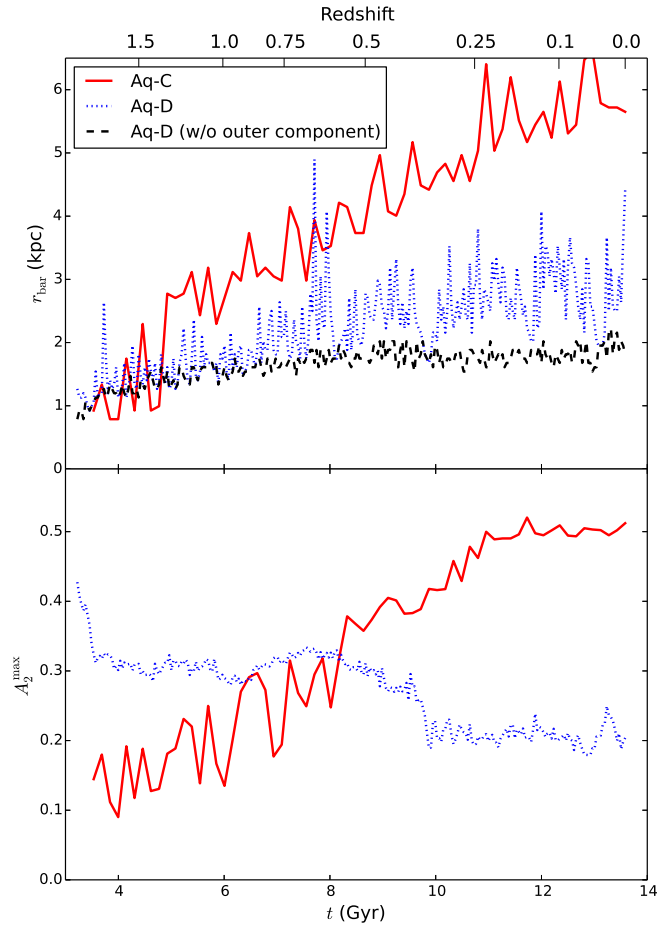} 
  \end{center}
\caption{
  Time evolution of the bar length, $r_\mathrm{bar}$, and the amplitude, 
  $A_2^\mathrm{max}$. 
  The upper panel shows the bar lengths as functions of time (or redshift). 
  The red solid and blue dotted lines indicate Aq-C and Aq-D, respectively. 
  The black dashed line represents the bar length of Aq-D when we exclude 
  the outer component.  
  The amplitude, $A_2^\mathrm{max}$, is shown in the lower panel. 
}
\label{fig:zevo}
\end{figure}

We now investigate the redshift evolution of the bars. 
We plot the bar lengths, $r_\mathrm{bar}$, and the bar amplitudes, 
$A_2^\mathrm{max}$, as functions of cosmic time and redshift in 
figure~\ref{fig:zevo}. 
We find that Aq-C's bar becomes longer and stronger with time.
This behavior is usually seen for bars in early-type 
disk galaxies \citep{combes93} or bars in galaxies with centrally  
concentrated dark halos in the idealized 
simulations \citep{athanassoula03}. 
This is consistent with the fact that Aq-C has a massive bulge 
\citep{okamoto13} and its halo shows the high central concentration 
(figure~\ref{fig:vc}). 

Contrarily, Aq-D's bar does not show such strong evolution in 
the length and its amplitude is almost constant until $t \sim 9.8$~Gyr, 
around which it sharply drops to the lower value. 
We also compute the bar length in Aq-D by excluding the 
component corresponding to the outer peak. 
To do so, we find the radius between the inner and outer peaks 
at which the amplitude, $A_2(r)$, takes the minimum value. 
If this radius is shorter than the bar length defined by the 
phase profile, we employ this radius as the bar length. 
The bar length defined this way exhibits very weak time evolution. 
The comparison between two bar lengths with the different definitions 
implies that 
the outer region of Aq-D's bar is violently disturbed  
by the interactions with the stellar clumps. 
We will show both bar lengths when the bar length is matter.  

\subsection{Pattern speed of the bars} \label{sec:pattern}

\begin{figure}
  \begin{center}
  \includegraphics[width=\linewidth]{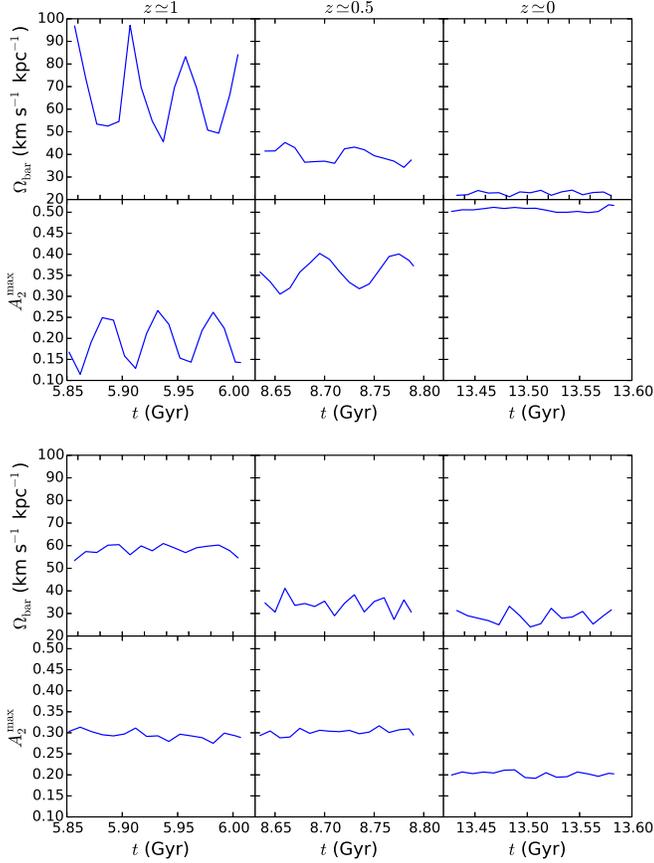} 
  \end{center}
\caption{
  Time evolution of the bar pattern speed, $\Omega_\mathrm{bar}$, 
  and the amplitude, $A_2^\mathrm{max}$, around $z = 1$, 0.5, and 0 
  (from left to right). 
  We show Aq-C and Aq-D in the upper and lower six panels, respectively. 
  The bar pattern speed and the amplitude are respectively shown 
  in the upper and lower three panels of each group of panels 
  as functions of cosmic time. 
}
\label{fig:pattern}
\end{figure}
\begin{figure}
  \begin{center}
  \includegraphics[width=\linewidth]{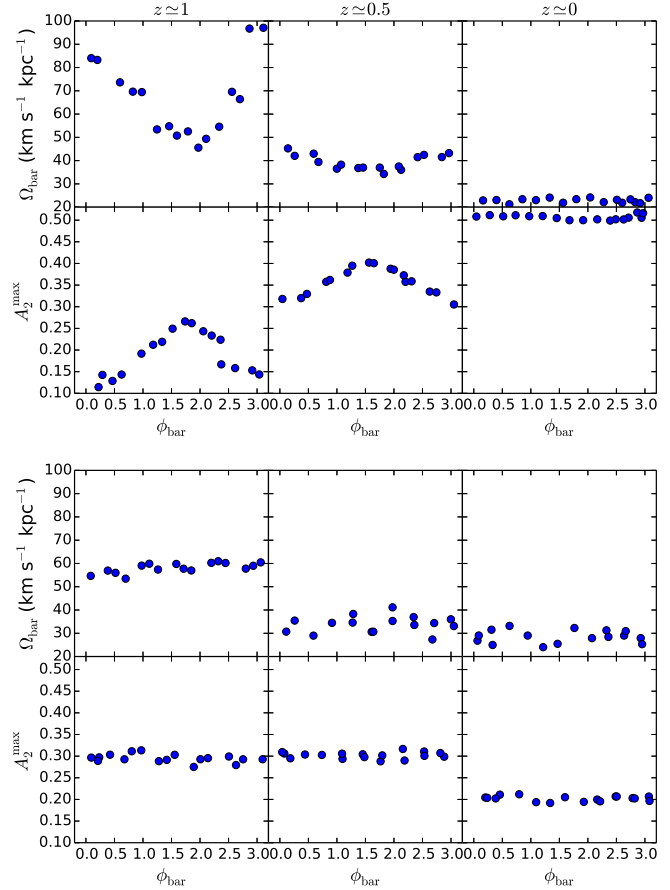} 
  \end{center}
\caption{
  Same as figure~\ref{fig:pattern} but now as functions of the 
  bar angle, $\phi_\mathrm{bar}$. 
}
\label{fig:patternphi}
\end{figure}

The time-steps used to output the simulation snapshots are
too large to measure the bar pattern speed. 
Moreover, the change of the disk orientation with time \citep{okamoto13} 
makes it difficult to measure the pattern speed. 
We hence restart the simulations around $z = 1$, 0.5, and 0, and then 
follow their evolution for a short period of time during which 
the orientation of the disks hardly changes; 
when the rotation axis of the disk is most misaligned with the initial one, the cosine of the angle between them is $0.996$, 
which occurs in the simulation for Aq-C at $z \sim 1$. 
In other simulations this value is always greater than $0.999$. 
The disk does not tumble during these periods, and thus the misalignment 
monotonically increases with time.  
We employ the output time-step of $10$~Myr for these additional simulations. 

In figure~\ref{fig:pattern} we show the bar pattern speed, 
$\Omega_\mathrm{bar} \equiv \dot{\phi}_\mathrm{bar}$, and the amplitude, 
$A_2^\mathrm{max}$, as functions of time around $z = 1$, 0.5 and 0. 
Firstly, we find that the pattern speed of the bars slows down 
from $z = 1$ to 0 in both galaxies. 
The long-term behavior of Aq-C's bar is consistent with what we expect 
from the results of the idealized simulations, that is, the bar becomes 
stronger and longer as it rotates more slowly. 
Aq-D's bar however does not change its amplitude between  
$z \simeq 1$ and 0.5 while it's pattern speed decreases.  
the bar amplitude becomes smaller from $z \simeq 0.5$ to $0$.

The pattern speed and the amplitude of Aq-C's bar display large 
oscillations at $z \simeq 1$ and 0.5. Interestingly, the 
short-term behavior is similar to the long-term one, i.e. the bar gets 
stronger when the pattern speed decreases. This oscillation 
becomes smaller with time and almost vanishes at $z \simeq 0$. 
Such oscillations have also been observed in idealized simulations 
\citep{dubinski09, amr13}.  
\citet{amr13} suggest that the oscillation is caused by the 
interaction between a bar and a host triaxial halo. 
We will investigate halo properties in section~\ref{sec:halo}. 

The small oscillations seen in the pattern speed of Aq-D's bar do not 
correlate with the oscillations of the bar amplitude; the amplitude is 
more or less constant during each period of time.  
The time evolution of the direction of the rotation axis is not the 
reason for the oscillations because the angle with the initial 
rotation axis ($z$-direction) is sufficiently small. 
Moreover this angle monotonically increases with time and does not 
oscillate.  
We suspect that the oscillation of the pattern speed of Aq-D's bar is 
caused by the interactions between the bar and the clumps. 

To see the short-term behaviors more closely, we now plot 
the pattern speed and the amplitude as functions of the bar angle, 
$\phi_\mathrm{bar}(t)$, in figure~\ref{fig:patternphi}.  
We find that the periods of the oscillations in the amplitude 
and the pattern speed are half a revolution period of the bar 
when they show large oscillations (at $z \simeq 1$ and $0.5$ in Aq-C). 
On the other hand, we do not see such a periodicity for Aq-D's bar. 

\subsection{Which component obtains the angular momentum?} \label{sec:torque}

\begin{figure*}
  \begin{center}
  \includegraphics[width=\linewidth]{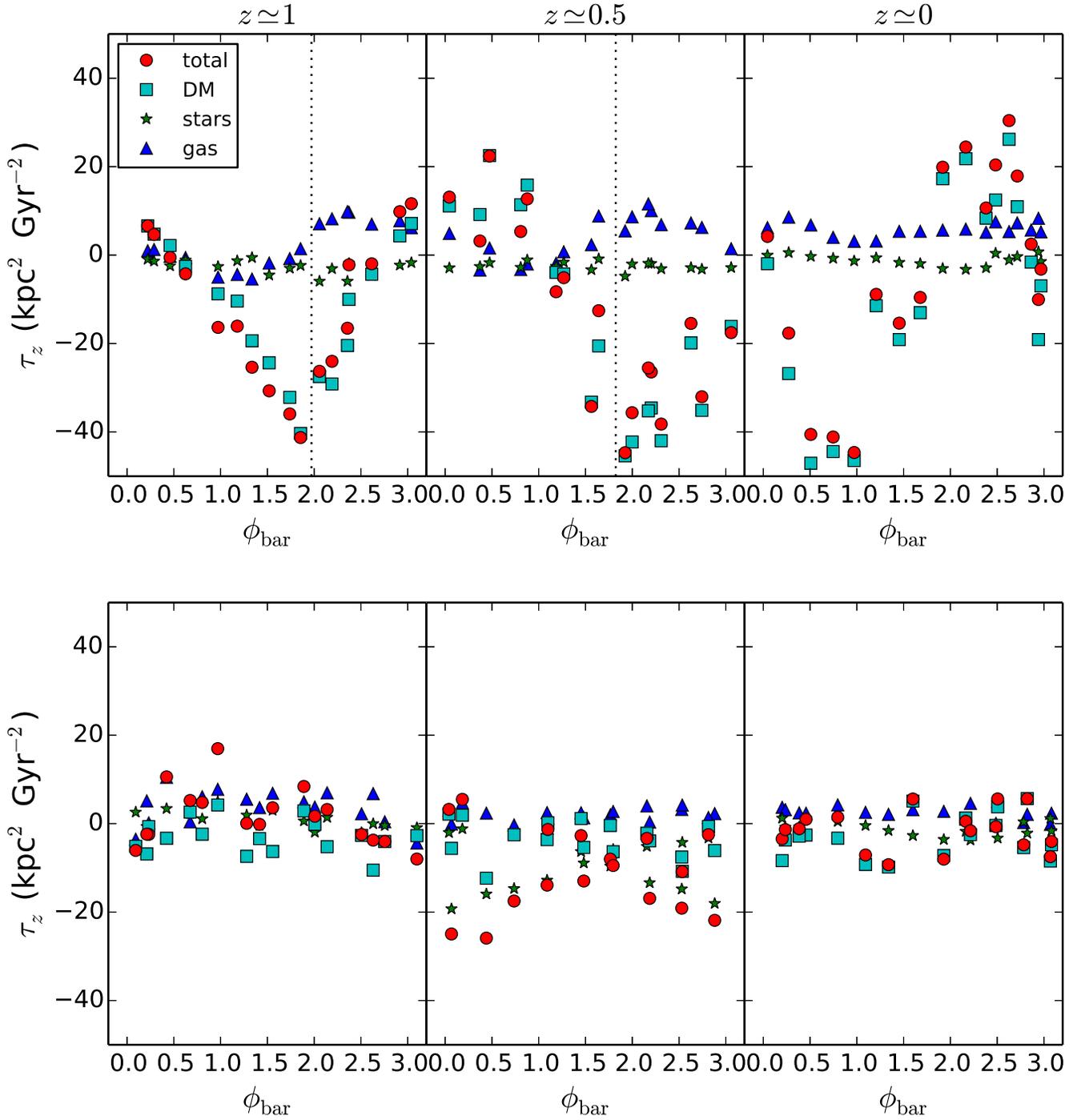} 
  \end{center}
\caption{
  The $z$-component of the specific gravitational torque acting on 
  the stars in the bar regions at $z \simeq 1$, 0.5, and 0. 
  The upper and lower panels show Aq-C and Aq-D, respectively. 
  The red circles indicate torque from all the particles 
  within virial radius, $r_\mathrm{vir}$. 
  The cyan squares, green stars, and blue diamonds respectively 
  show the contributions of the dark matter, stars, and gas. 
  The vertical dotted lines in the upper left and the upper middle 
  panels indicate the bar angle at which the bar pattern speed is 
  the smallest.  
}
\label{fig:torque}
\end{figure*}
\begin{figure*}
  \begin{center}
  \includegraphics[width=\linewidth]{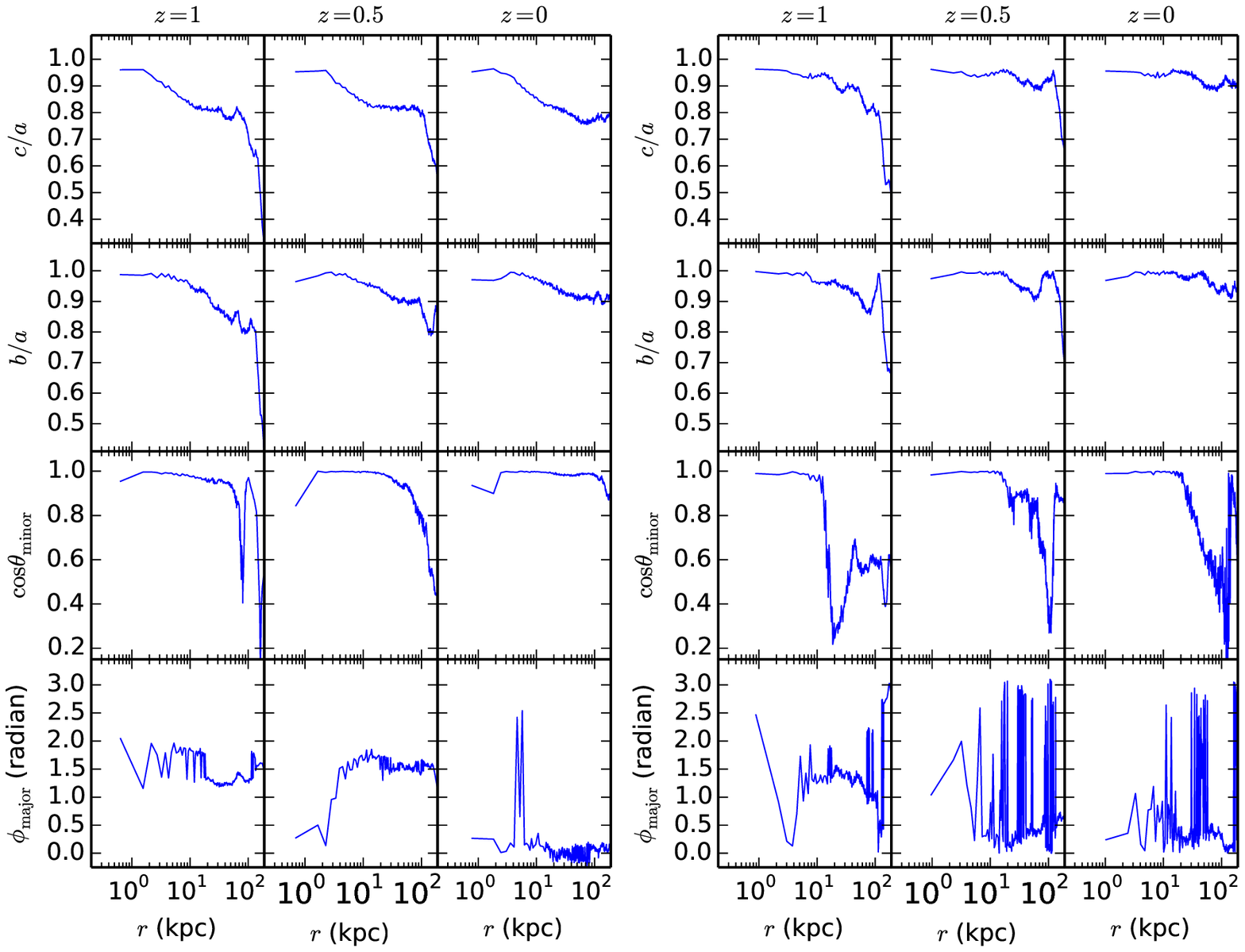} 
  \end{center}
\caption{
  Radial profiles of the halo triaxiality. 
  The left panels show the halo properties of Aq-C and the right 
  ones show those of Aq-D. 
  From top to bottom, we display the radial profiles of the 
  minor-to-major axial ratio, $c/a$, the intermediate-to-major 
  axial ratio, $b/a$, the directional between the disc 
  ration axis and the halo minor axis, and the azimuthal direction 
  of the halo major axis. 
  From left to right, we show the halos at $z = 1$, $0.5$, and $0$ 
  for each galaxy. 
}
\label{fig:shape_diff}
\end{figure*}

Since the bar pattern speed slows down from $z = 1$ to 0, 
we expect that the angular momentum of the bars is transferred to 
other components, such as outer disks or dark halos. 
In order to identify which component plays the most important 
role in spinning down the bars, we try to measure the torque 
from each component acting on the bars. 
Doing this is however not straightforward because defining 
particles that constitute the bar is not a simple task.  
It is also difficult to relate the measured torque to the 
change in the bar pattern speed even if we somehow define 
the particles that belong to the bar, since
the angular momentum of the bar, 
$L_\mathrm{bar} = I_\mathrm{bar} \Omega_\mathrm{bar}$, 
where $I_\mathrm{bar}$ is the moment of the inertia of the bar, 
is different from the total angular momentum of the particles
that constitute the bar.

We therefore take a simpler approach. 
We define the bar region as a disk with the radius, $r_\mathrm{bar}$,  
and the height 1~$h^{-1}$~kpc, and then compute the torques acting on 
the star particles within this region. 
Since the $z$-component of the torque acting on an axisymmetric
component is zero, the torque mainly operates on the bar as long as the bar 
is the most significant non-axisymmetric structure in the region. 
We have checked that we obtain qualitatively equivalent results when we 
calculate torques on the particles along the bar. 
We have also confirmed the results does not change qualitatively if we employ 
the bar length by excluding the outer $m = 2$ component for Aq-D. 

In figure~\ref{fig:torque}, we show the $z$-component of the specific 
torque from the particles within the virial radius acting on the stars 
in the bar region as a function of the bar angle, $\phi_\mathrm{bar}(t)$.  
We find that the torque from the dark halo dominates the total 
torque in Aq-C and it is negative on average; the torque from the dark 
matter spins down the stars in the bar region more strongly at higher 
redshift.   
In Aq-C, the torque from the dark matter shows the periodic change with 
the period of half a bar revolution period. 
This behavior suggests that the torque is exerted by the anisotropic 
distribution of the dark matter. 

Although the period with which the torque changes is the same as the 
period with which the pattern speed changes in Aq-C at $z \simeq 1$ and 
$0.5$, the change in the angular momentum of the stars within  
the bar region is not directly reflected to the bar pattern speed. 
In the panels of $z \simeq 1$ and $0.5$ for Aq-C, we indicate the
bar angle where the bar pattern speed becomes the smallest. 
We find that, against the intuition, the torque takes the largest 
negative value where the bar pattern speed becomes the smallest. 
The pattern speed should be the smallest where the torque changes its 
sign from minus to plus  
if the angular momentum of the stars within the bar region were 
directly reflected to the bar pattern speed. 
Moreover, the pattern speed of Aq-C's bar at $z \simeq 0$ is almost 
constant while the torque show the same level of variation as at
$z \simeq 1$ and 0. 
This is a direct consequence of the fact that the bar angular momentum, 
$I_\mathrm{bar} \Omega_\mathrm{bar}$, is different physical quantity 
from the total angular momentum of the particles belonging to the 
bar as we mentioned earlier. 

In Aq-D, the amplitude of the specific torque is much smaller than 
that in Aq-C. The long-term evolution of the bar pattern speed in 
Aq-D is thus much more moderate than Aq-C. 
The torque is not always dominated by the contribution of the 
dark matter, for example the torque is dominated by the contribution  
of the stars at $z \simeq 0.5$. 
This is consistent with the fact that the central density of the dark 
halo of Aq-D is much lower than that of Aq-C and thus there is less 
dark matter to absorb the angular momentum of the bar. 

\citet{athanassoula03} show that the angular momentum is emitted 
from near-resonant material at the inner Lindblad resonance (ILR), 
i.e. a bar, and absorbed by mainly by near-resonant material at the 
corotation and the outer Lindblad resonance in the halo; 
the near-resonant material in the outer disk also absorbs the angular 
momentum, but the role the outer disk plays is much less significant than 
the halo. 
This picture is in good agreement with the analytic calculations 
\citep{lynden-bell72, tremaine84}. 
Our results show that the halo is the main absorber of the angular momentum 
when a significant amount of the angular momentum of the bar is transferred 
from the bar. The results thus qualitatively agree with those by the idealized 
simulations and the prediction of the analytic calculations.  

\subsection{Halo properties} \label{sec:halo}

\citet{amr13} claim that the oscillations seen in the bar amplitude can 
be understood by the interaction between a bar and its host triaxial halo. 
They find that $A_2^\mathrm{max}$ has a minimum and the halo 
intermediate-to-major axial ratio, $b/a$, has a maximum when 
the bar and halo major axis are aligned. On the other hand, 
$A_2^\mathrm{max}$ has a maximum and the halo $b/a$ has a minimum 
when the bar and halo major axis are perpendicular.  
These results imply that the triaxial halo plays a role of an outer bar 
and the basic building blocks of the bar are loops.

We thus investigate the halo triaxiality in this section. 
We have removed the subhalos identified by {\scriptsize SUBFIND} 
\citep{spr01} in order to consider only the smooth component of 
the dark halos. 
The results are however almost identical to the case in which 
we use the dark matter distribution as it is, except at the 
outer parts of the dark halos.  

In figure~\ref{fig:shape_diff}, we show the radial profiles of
the halo triaxiality, where each radial bin contains 20000 dark 
matter particles. 
The inner halo has a higher  sphericity than the outer halo and 
the inner halo's  minor axis is well aligned with the disk rotation axis. 
The direction of the minor axis of the outer halo is almost independent 
from that of the inner halo. 
The baryonic contraction due to gas cooling makes the inner halo 
more spherical than dark matter only simulations \citep{kazantzidis04, bai05}. 
The tidal torque between the disk and the inner halo aligns the rotation 
axis and the halo minor axis \citep{bai05}. 
We also find that Aq-D's halo is more spherical than Aq-C's. 
This explains the small amplitude of the specific torque from 
dark matter acting on the bar in Aq-D.  

%
\begin{figure}
  \begin{center}
  \includegraphics[width=\linewidth]{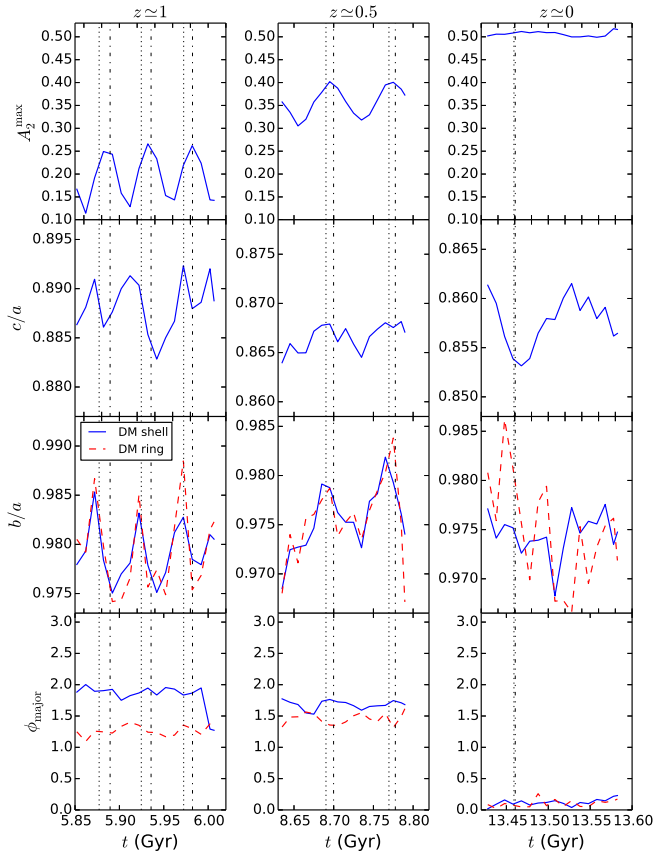} 
  \end{center}
\caption{
  Time evolutions of the bar amplitude and the halo triaxiality of Aq-C 
  around $z = 1$, $0.5$ and $0$ (from left to right). 
  From top to bottom, we present the bar amplitude, $A_2^\mathrm{max}$, 
  the minor-to-major halo axial ratio, $c/a$, the intermediate-to-major 
  halo axial ratio, $b/a$, and the azimuthal direction of the halo major 
  axis, $\phi_\mathrm{major}$. 
  We also show the axis ratio of the dark matter ring in the disk plane and 
  the azimuthal direction of its major axis by the red dashed lines. 
  The vertical dashed lines indicate the time when the directions of the halo
  major axis and the bar are aligned and the vertical dotted lines indicate 
  the time when the directions of the major axis of the dark matter ring and bar  
  are aligned. 
}
\label{fig:halo_oscillation}
\end{figure}

We now explore the time oscillations of the halo triaxiality of Aq-C that 
shows oscillations in the bar amplitude and pattern speed at high redshift. 
\citet{am02} claim that the halo triaxiality should be measured 
in a density bin instead of a radial bin since the former is noisier 
than the latter. 
It is however not the case in our simulations probably because 
the dark matter density distribution in cosmological halos are not as 
smooth as that in the idealized simulations even if we remove 
the subhalos. 
We hence measure the triaxiality of a spherical shell of radius 
between $r_\mathrm{bar}$ and $2 r_\mathrm{bar}$, where we employ 
the mean bar length during each simulation as $r_\mathrm{bar}$. 
We first sort the dark matter particles by the distance from the center
and identify the ranking of the particles that lie between 
$r_\mathrm{bar}$ and $2 r_\mathrm{bar}$ in the first snapshot. 
In the rest of the snapshots, we use the dark matter particles 
that have the same ranking in radius as the particles selected in the 
first snapshot to measure the orientation and the shape of the halo. 
Our results are not sensitive to the choice of the radius of the shell 
as long as that lies between $r_\mathrm{bar}$ and $3 r_\mathrm{bar}$. 
Within $r_\mathrm{bar}$, the direction of the halo major axis largely 
oscillates, but not rotates. In the outer part,  
($r \gtrsim 3 r_\mathrm{bar}$), the direction of the major axis of the 
halo is independent of that of the inner part 
(figure~\ref{fig:shape_diff}). 
Since the minor axis of the halo is not parallel to the rotation axis 
of the disk as shown in figure~\ref{fig:shape_diff}, the azimuthal 
direction of the major axis of the halo can be different from that of the 
major axis of the dark matter distribution in the disk plane as we will 
show later. 
We thus measure the axis ratio of the dark matter ring of $r_\mathrm{bar} < 
r < 2 r_\mathrm{bar}$ and $|z| < 0.1 h^{-1}$~kpc and the direction of its 
major axis, too. 

In figure~\ref{fig:halo_oscillation}, we show the time evolution of the bar 
amplitude, the minor-to-major halo axial ratio,  
and the intermediate-to-major halo axial ratio,
together with the axial ratio of the dark matter ring. 
We also show the azimuthal directions of the major axes of the dark matter 
shell and ring. The directions of these axes are better aligned at lower 
redshift simply because the directions of the halo minor axis and the disk 
rotation axis are better aligned at lower redshift as shown in 
figure~\ref{fig:shape_diff}. 

We find that the axial ratios, $b/a$, of the shell and ring oscillate with 
the same frequency as the bar oscillation. 
The minor-to-major axial ratio, $c/a$, also shows similar oscillations to 
$b/a$. 
These results agree with the finding by \citet{amr13}.  

The bar amplitude is however largest when the bar and the major axis 
of the dark matter ring are parallel. This behavior contradicts the 
result by \citet{amr13} and the loop concept that predicts a loop 
corresponding to an inner bar is less elongated when the inner 
and outer bars are aligned \citep{maciejewski00, maciejewski07}.  
The simulation results by \citet{heller01} and \citet{heller07b}
are more analogous to ours. They find that the inner bar/ring component 
is more elongated when it is parallel to the outer bar. 
On the other hand, the halo is more axisymmetric, i.e. $c/a$ and $b/a$ are 
large, when the bar and the major axis of the dark matter ring are aligned. 
This result is consistent with the behaviors of the halo triaxiality 
found by \citet{amr13}. 
Further studies are needed to understand the origin of the  
disagreement among the simulations. 

\subsection{Interaction with the $m = 4$ Fourier mode} \label{sec:m4}

\begin{figure}
  \begin{center}
  \includegraphics[width=\linewidth]{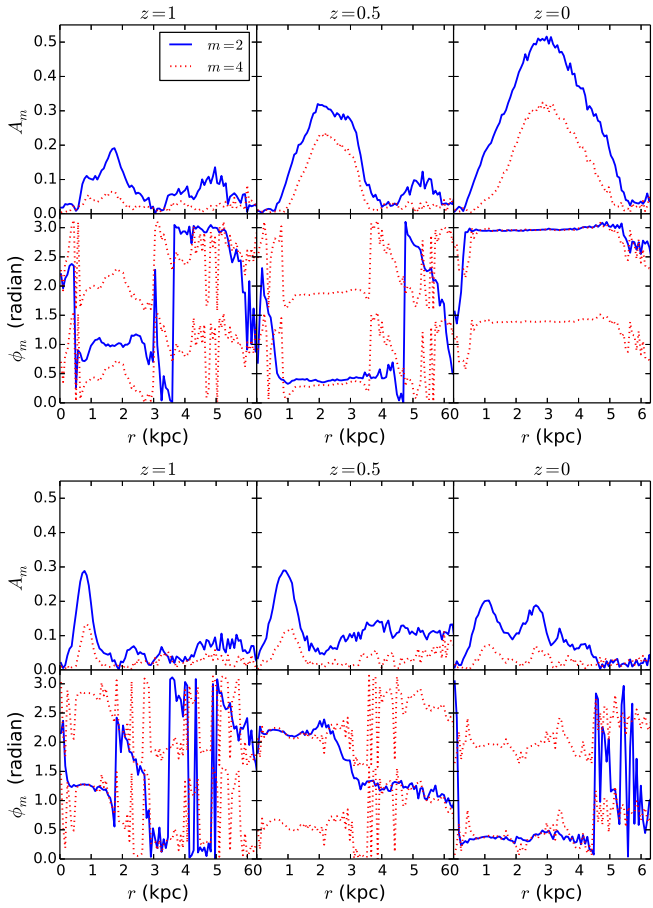} 
  \end{center}
\caption{The amplitude and phase profiles of the $m = 2$ and $m = 4$ modes. 
  The upper and lower six panels show Aq-C and Aq-D, respectively. 
  The blue solid and red dotted lines respectively represent the 
  $m = 2$ and $m = 4$ modes. 
  The amplitude and phase are shown in the upper and lower panels 
  of each group of panels. 
  From left to right, the results at $z = 1$, 0.5, and 0 are presented.  
  For the phase, $\phi_4(r)$, we also plot $\phi_4(r) + \pi/2$ in 
  order to compare it with the $m = 2$ mode. 
}
\label{fig:m4}
\end{figure}

In this subsection, we investigate the second most significant Fourier 
mode, the $m = 4$ mode, and explore possible interaction between 
the $m = 2$  and $m = 4$ Fourier modes. 
In figure~\ref{fig:m4}, we plot the amplitude and the phase profiles of 
the $m = 4$ Fourier mode. Since the phase of the $m = 4$ mode, 
$\phi_4(r)$, is defined between 0 and $\pi/2$, 
we also plot $\phi_4(r) + \pi/2$ to investigate the alignment between $m = 4$ and $2$ modes. Note that as Aq-C's snapshot at $z \simeq 1$, we chose 
the one in which the misalignment between the two components is evident. 

%

We find that the relative importance of the $m = 4$ component to the 
$m = 2$ component is larger in Aq-C than in Aq-D. 
This result is consistent with those obtained by many simulations. 
In \citet{am02}, the relative importance of the higher order even moments 
is higher in a galaxy with a stronger bar. 
The same trend is seen in cosmological bars \citep{scannapieco12}. 
In Aq-C, the maxima of $A_4(r)$ occur at almost the same radii as 
the maxima of $A_2(r)$, i.e. at $r_2^\mathrm{max}$. 
Usually in idealized simulations, the maxima of the amplitudes of the higher 
order even moments occur considerably larger radii than $r_2^\mathrm{max}$ 
in galaxies with massive halos \citep{am02, athanassoula03}.  
\citet{scannapieco12} obtain the same result as ours for the same Aquarius 
halo, Aq-C, in spite of their lower resolution and weaker feedback, 
i.e. a heavier disk, than ours \citep{sca09}. 
This agreement implies that the halo formation history and its shape play 
an important role in shaping substructure of a galaxy.  


In Aq-C, the $m = 2$ component is largely misaligned with the $m = 4$ 
component  at $z = 1$ and is slightly misaligned at $z = 0.5$. 
At $z = 0$ the two modes are perfectly aligned with each other. 
On the other hand, the two modes are always aligned in Aq-D. 
It is interesting that the bar and the $m = 4$ component are 
misaligned when the bar shows large oscillations in its pattern speed 
and amplitude. 

Next, we explore the relation between the pattern speed and amplitude 
of the $m = 2$ and $m = 4$ components. 
To calculate the pattern speed of the $m = 4$ component, 
we define $A_4^\mathrm{max}$ and $r_4^\mathrm{max}$ by exactly the same way 
as we defined $A_2^\mathrm{max}$ and $r_2^\mathrm{max}$ and then we
define the phase of the $m = 4$ component as $\phi_4(r_4^\mathrm{max})$. 

\begin{figure}
  \begin{center}
  \includegraphics[width=\linewidth]{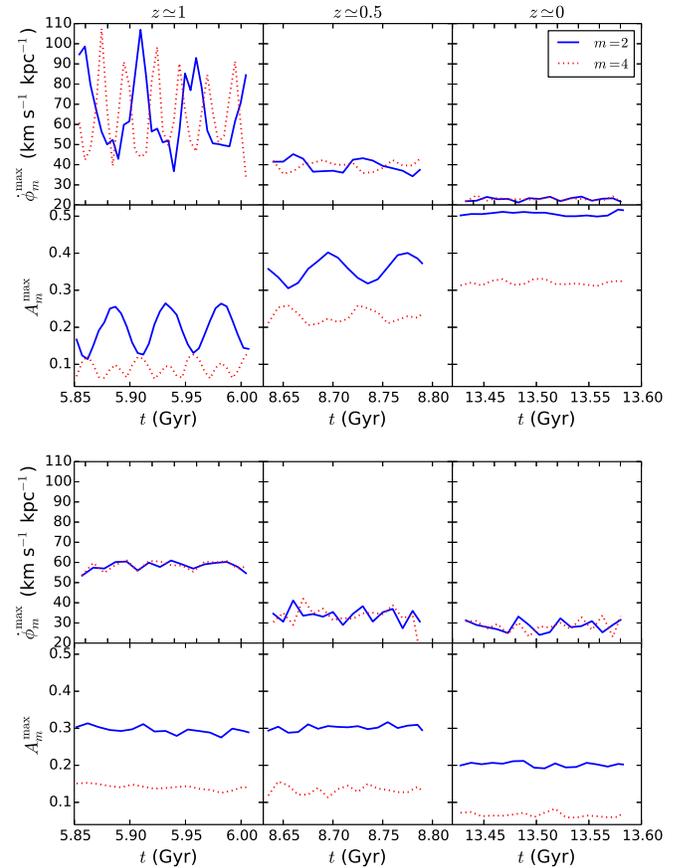} 
  \end{center}
\caption{
  Same as figure~\ref{fig:pattern} but we now show the pattern speed 
  and amplitude of the $m = 4$ components as well. 
  The $m = 2$ and $m = 4$ components are respectively represented by 
  the blue solid and red dotted lines.  
}
\label{fig:m4pattern}
\end{figure}

In figure~\ref{fig:m4pattern}, we show the pattern speed of the $m = 2$ 
component, $\Omega_\mathrm{bar} \equiv \dot{\phi}_\mathrm{bar}$, and 
the $m = 4$ component, $\dot{\phi}_4(r_4^\mathrm{max})$. 
We find that the pattern speed of the $m = 4$ component also shows large and 
periodic oscillation at $z \simeq 1$ in Aq-C. 
This oscillation becomes much smaller at $z \simeq 0.5$ at which 
the misalignment between the two modes is small (see figure~\ref{fig:m4}). 
Once the phases of the two components are perfectly aligned with each other, 
the pattern speed of the both components becomes almost constant and of 
course the two components have the same pattern speed. 

The amplitude of the $m = 4$ component also varies with the same frequency 
as the pattern speed. As for the $m = 2$ component, the amplitude becomes 
large when the pattern speed is small and vise versa. 
The frequency of the oscillation of the pattern speed of the $m = 4$ 
component is clearly higher than that of the $m = 2$ component. 
We find that the period of the oscillations of 
the pattern speed and the amplitude of the $m = 4$ component is 
a quarter of its revolution period. 

As we have shown, the $m = 2$ and $m = 4$ components in Aq-D are always 
aligned with each other. In this case, the pattern speed and the amplitude of 
both $m = 2$ and $m = 4$ components do not show the periodic oscillations, 
which are seen in Aq-C at $z \simeq 1$ and 0.5. 
The high-frequency oscillations are probably caused by the interactions 
with clumps as we have already discussed for the $m = 2$ mode.  
Our results suggest that the pattern speed of the bar oscillates if its 
phase is misaligned with the phase of the $m = 4$ component in the same 
region. 

\subsection{Resonances and the bar length} 

\begin{figure}
  \begin{center}
  \includegraphics[width=\linewidth]{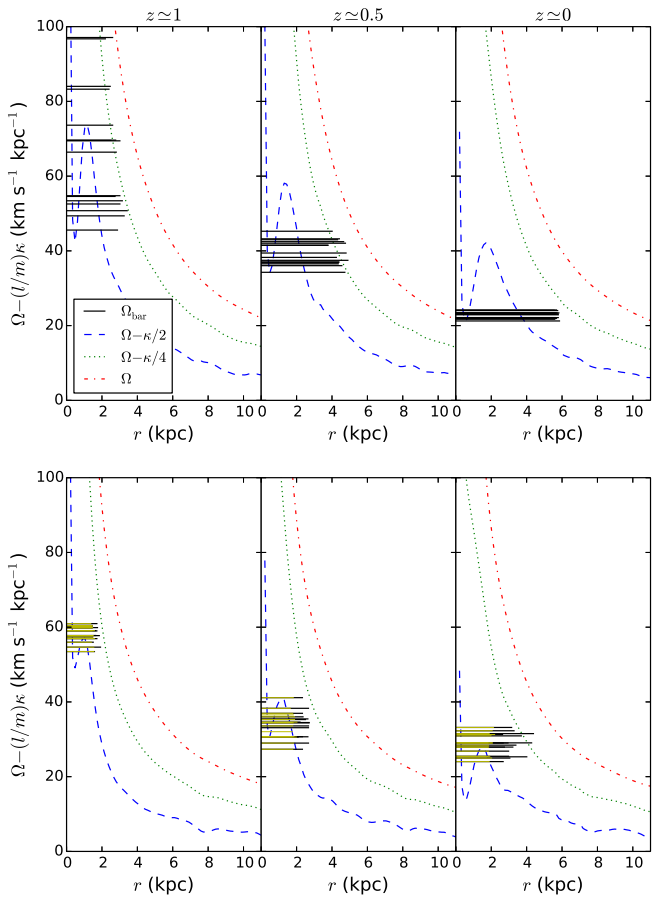} 
  \end{center}
\caption{
  Behavior of $\Omega - \kappa/2$, $\Omega - \kappa /4$ and $\Omega$ 
  at $z = 1$, 0.5, and 0.  
  The blue dashed, green dotted, and red dot-dashed lines respectively 
  represent $\Omega - \kappa/2$, $\Omega - \kappa/4$, and $\Omega$. 
  The upper and lower panels indicate Aq-C and Aq-D, respectively, and 
  redshifts are $1$, $0.5$, and $0$ from left to right. 
  We show the bar pattern speed around these redshifts by the horizontal 
  black solid lines, whose lengths correspond to the bar lengths. 
  For Aq-D, we also show the case in which we define the bar lengths 
  by excluding the outer component (horizontal yellow solid lines). 
}
\label{fig:resonance}
\end{figure}

Finally we investigate the resonance structure of the simulated galaxies. 
In figure~\ref{fig:resonance} we show the behaviors of $\Omega - \kappa/2$, 
$\Omega - \kappa/4$, and $\Omega$ as functions of radius,  
where $\Omega$ is the angular frequency of a circular orbit and $\kappa$ 
is the radial angular frequency. 

The sharp rise of the $\Omega - \kappa/2$ curves towards the center is 
due to the gravitational softening.  
Since the spatial extent of this region is much smaller than the scale 
we are interested in, we will ignore the innermost 2:1 resonance due to 
this sharp rise in the following discussion. 
Each $\Omega - \kappa/2$ curve has a peak and thus these galaxies 
have double ILRs if the bar pattern speed is smaller than the 
peak value (and if we ignore the inner most ILR). 
The peak values decrease with time since the central density of 
the galaxies decrease with time due to the stellar evolution and the 
feedback (see figure~\ref{fig:vc}). 

The density structure does not change on the short time-scale during 
which we measure the pattern speed of the bars around $z = 1$, 0.5, and 0. 
We show the bar pattern speed and the bar length measured around these 
redshifts by the horizontal lines.  
From the behavior of the bar pattern speed and the amplitude in Aq-C 
around $z = 1$ and 0.5, we speculate that  
the smaller the pattern speed is, the larger the amplitude is  
if the $\Omega - \kappa/2$ curve is fixed \citep{kormendy13}. 

On the long time-scale, $\Omega - \kappa/2$ curve is lowered as shown in 
figure~\ref{fig:resonance}.
In Aq-C, the bar pattern speed strongly decreases from $z = 1$ to 0. 
As a result, the bar amplitude becomes larger with time 
(figure~\ref{fig:zevo}). 
In Aq-D, the slow down rate of the bar pattern speed is much lower than that 
in Aq-C. Consequently, the peak value of $\Omega - \kappa/2$ curve decreases 
as fast as or faster than the bar pattern speed. 
The bar amplitude in this case does not change much or becomes smaller 
as shown in figure~\ref{fig:zevo}. 
At $z \simeq 0$, there is no 2:1 resonances in many cases if we ignore the 
inner most resonance.  
As a result, the bar amplitude at $z = 0$ is smaller than that at $z = 0.5$. 
We will discuss the sharp decline in the amplitude of Aq-D's bar at 
$t \sim 9.8$~Gyr shown in figure~\ref{fig:zevo} in the next section. 

The bar lengths are around the 4:1 resonances in both galaxies at 
all redshifts.  
Even if we define the length of Aq-D's bar by excluding the outer component, 
the bar looks terminated at this resonance except at $z \simeq 0$. 
\citet{patsis97} study the orbital structure in the potential model of 
NGC 4314 and predict that the longest stable periodic orbits are 
found at the 4:1 resonance. 
Our cosmological simulations confirm their prediction. 

\section{Discussion and conclusions}

\begin{figure}
  \begin{center}
  \includegraphics[width=\linewidth]{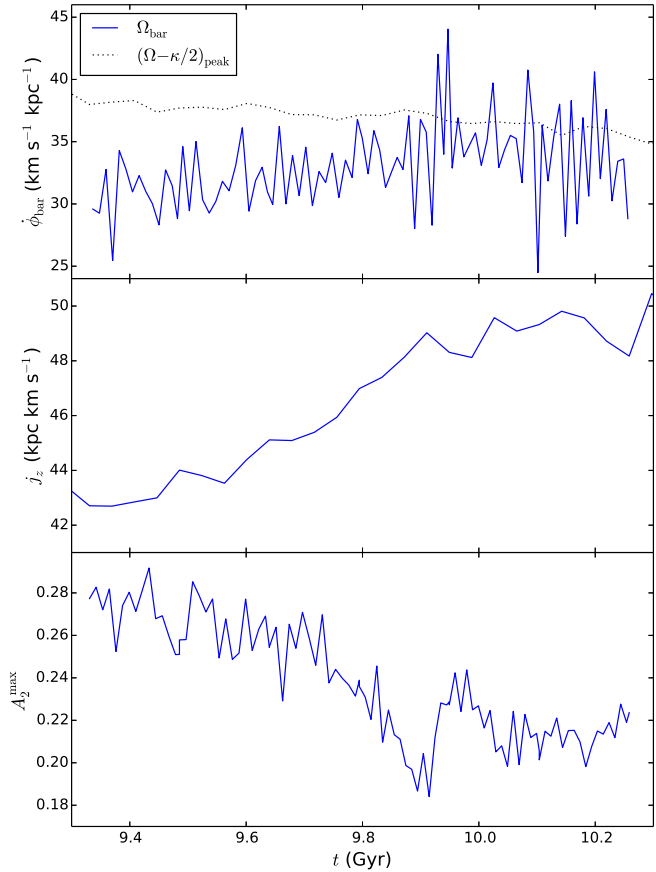} 
  \end{center}
\caption{
  Evolution of the bar pattern speed and amplitude of Aq-D 
  around $t \simeq 9.8 Gyr$.  
  In the top panel, we show the evolution of the bar pattern speed 
  and the peak value of the $\Omega - \kappa/2$ curve by the blue solid and
  black dotted lines respectively. 
  In the middle panel we also plot the specific angular momentum of the 
  stars in the bar region ($r < 1.68$~kpc and $|z| < 1~h^{-1}$~kpc).  
  In the bottom panel, we show the bar amplitude $A_2^\mathrm{max}(t)$. 
}
\label{fig:resoevo}
\end{figure}

We have investigated the cosmological evolution of bars by utilizing 
the two Milky Way-mass galaxies formed in the fully self-consistent 
simulations of the galaxy formation by \citet{okamoto13}. 
The evolution qualitatively agrees with what is expected from idealized 
simulations, that is, the bar in the galaxy having more centrally 
concentrated mass distribution exhibit stronger evolution than the 
other \citep{combes93, athanassoula03}.  
This picture is in good agreement with observations (e.g. \cite{cheung13}). 
The bar in Aq-C receives the large negative torque from the dark matter 
and is significantly slowed down $z = 1$ to 0. 
As the bar pattern speed decreases, the bar becomes stronger and longer. 
On the other hand, the torque acting on Aq-D's bar is much weaker by 
lacking materials that absorb the angular momentum of the bar.

The pattern speed of Aq-C's bar violently oscillates at $z \simeq 1$. 
The oscillation becomes smaller at $z \simeq 0.5$ and disappears at 
$z \simeq 0$. The period of the oscillation as a function 
of the bar angle is $\pi$, indicating the interaction with the halo's 
quadrapole moment.
The amplitude of the bar also shows the oscillations with the same frequency 
as those of the pattern speed. 
When the bar rotates slower, it becomes stronger and vise versa. 
This short-term behavior is consistent with the long-term behavior in Aq-C, 
i.e. the slower the bar rotation is, the lager its amplitude is.  
These oscillations correlate with the oscillations in the halo triaxiality 
as pointed out by \citet{amr13}. When the bar and the halo major axis 
are parallel, the halo is more spherical and the bar is strongest. 
The former result is agree with the simulations by \citet{amr13}, but 
the latter is opposite to them. The origin of the disagreement is 
unclear. 

The evolution of the bar in Aq-D seems to contradict this scenario. 
While the pattern speed of the bar decreases with time, its amplitude 
does not increase from $z = 1$ to 0.5 and even decreases from 
$z = 0.5$ to 0. 
We speculate that this is because the $\Omega - \kappa/2$ curve is 
lowered with time (figure~\ref{fig:resonance}) due to the decreasing 
central density with time (figure~\ref{fig:vc}), 
which is most likely caused by the mass loss and feedback from the 
stellar populations. 
The inclusion of the mass loss through entire life of stellar 
populations and strong feedback 
is one of the biggest differences between our cosmological simulations 
and idealized simulations of isolated galaxies. 
The instantaneous recycling approximation is often used in 
idealized simulations and old stars do not lose their mass even 
when gas cooling, star formation, 
and feedback are included in idealized simulations (e.g. \cite{amr13}). 
Thanks to the efficient angular momentum transfer from the bar to the 
dark matter in Aq-C, the bar pattern speed drops faster than 
the $\Omega - \kappa/2$ curve, and thus the bar becomes stronger. 
On the other hand, the slowdown rate of Aq-D's bar is as low as 
the decreasing rate of the peak value of the $\Omega - \kappa/2$ curve. 

In Aq-D, the amplitude of the bar sharply drops at $t \sim 9.8$~Gyr. 
To see why the bar is weakened, we measure the bar pattern speed 
and the peak value of the $\Omega - \kappa/2$ curve, 
$(\Omega - \kappa/2)_\mathrm{peak}$,  around this epoch.  
In figure~\ref{fig:resoevo}, we show the evolution of the bar pattern speed 
by restarting the simulations from several snapshots and compare it with 
the evolution of $(\Omega - \kappa/2)_\mathrm{peak}$. 

We find that the bar pattern speed is in fact an increasing function of time 
on average until $t \simeq 9.9$~Gyr, while 
$(\Omega - \kappa/2)_\mathrm{peak}$ slowly decreases with time. 
As the bar pattern speed approaches  $(\Omega - \kappa/2)_\mathrm{peak}$, 
the bar amplitude decreases. 
We also show the evolution of the specific angular momentum of the stars 
in the bar region ($r < 1.68$~kpc). 
The specific angular momentum in the bar region sharply increases with 
time until $t \simeq 9.9$~Gyr. 
The angular momentum is most presumably brought by the clumps. 
\citet{okamoto13} has shown that a non-negligible amount of mass is added to 
Aq-D's bulge by the clumps during this epoch.
The high frequency oscillations in the bar pattern speed is likely to be 
cause by the interactions with the clumps. 
At $t \gtrsim 9.9$~Gyr, the bar pattern speed often exceeds 
$(\Omega - \kappa/2)_\mathrm{peak}$. 
These results support the idea that the bar amplitude is determined by 
the relation between the pattern speed and the $\Omega - \kappa/2$ curve. 

Interestingly, the oscillations in the bar pattern speed and amplitude 
in Aq-D's bar are observed only when the $m = 2$ component is misaligned 
with the $m = 4$ component. 
The oscillations become smaller as the two components get aligned with 
each other. 
The $m = 4$ components have the comparable spatial size to the 
$m = 2$ components in both galaxies. 

The bar lengths seem to be determined by the 4:1 resonances.  
Our cosmological simulations thus confirm the prediction by 
\citet{patsis97}.  
Our results are also consistent with the observation of 
NGC~253, which has a bar whose length coincides 
with 4:1 resonance \citep{sorai00}. 

In summary, using high-resolution cosmological simulations of disk galaxy formation, we show that the bar evolution in the cosmological simulations is qualitatively consistent with that obtained by idealized simulations of isolated disk galaxies and observations. 
We find that the strong feedback and continuous mass loss from stellar populations significantly lowers the central density with time and hence changes the resonance structure. We also find that the clumps formed in the disk can spin up the bar and can weaken its strength.
Our sample is too small to understand why the oscillations in bar amplitude 
correlate with those in halo triaxiality differently from 
the idealized simulations by \citet{amr13} and thus we leave it for future 
studies. 

\bigskip
We are grateful to the anonymous referee for the careful reading of 
the manuscript and the thoughtful comments. 
We would like to thank Masafumi Noguchi and Junich Baba for helpful 
discussion. 
We also thank John  Kormendy,  Francoise Combes, and Shunsuke Hozumi  
for useful comments on the manuscript. 
Numerical simulations were carried out with Cray XC30 in CfCA at NAOJ 
and T2K-Tsukuba in Center for Computational Sciences at University 
of Tsukuba.
TO acknowledges the financial support of Japan Society for the Promotion of 
Science (JSPS) Grant-in-Aid for Young Scientists (B: 24740112). 

\appendix
\section{Clump removal} \label{app:clump}

\begin{figure}
  \begin{center}
  \includegraphics[width=\linewidth]{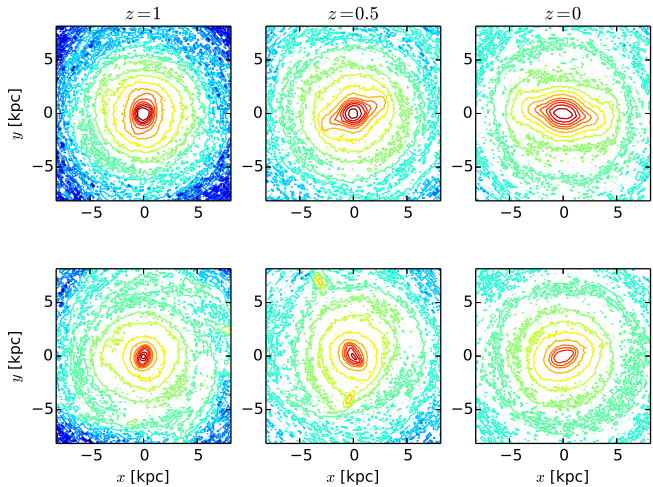} 
  \end{center}
\caption{
  The same as figure~\ref{fig:contour}, but shown the original 
  surface stellar density maps where the clumps are not removed). 
}
\label{fig:contour_orig}
\end{figure}

In order for the $m = 2$ component not to include contributions from clumps, 
we have removed the clumps from the density field before we calculate the 
Fourier series. 
In figure~\ref{fig:contour_orig}, we present the original surface stellar 
density maps. There are clearly two clumps in Aq-D at $z = 0.5$, which are 
not seen in figure~\ref{fig:contour} where we have eliminated the contribution 
of the clumps. 

To remove the clumps, we start from the cell that has the highest surface 
stellar density and then examine its neighboring cells. 
If the surface density of a neighboring cell is equal to or lower than that 
of the current cell, we add it to the {\it smooth} component. 
By recursively examining the neighboring cells of the smooth component, 
local density peaks are isolated as non-smooth components.   

We then replace the surface density of the cell at $(r, \phi)$,  
classified as the non-smooth components with that of the cell at 
$(r, \phi + \pi)$. This method has an advantage that it can be applied even 
when a clump is at radius comparable to or smaller than the bar length.  
We have applied this procedure to all the simulation snapshots used 
in this paper. 

\begin{figure}
  \begin{center}
  \includegraphics[width=\linewidth]{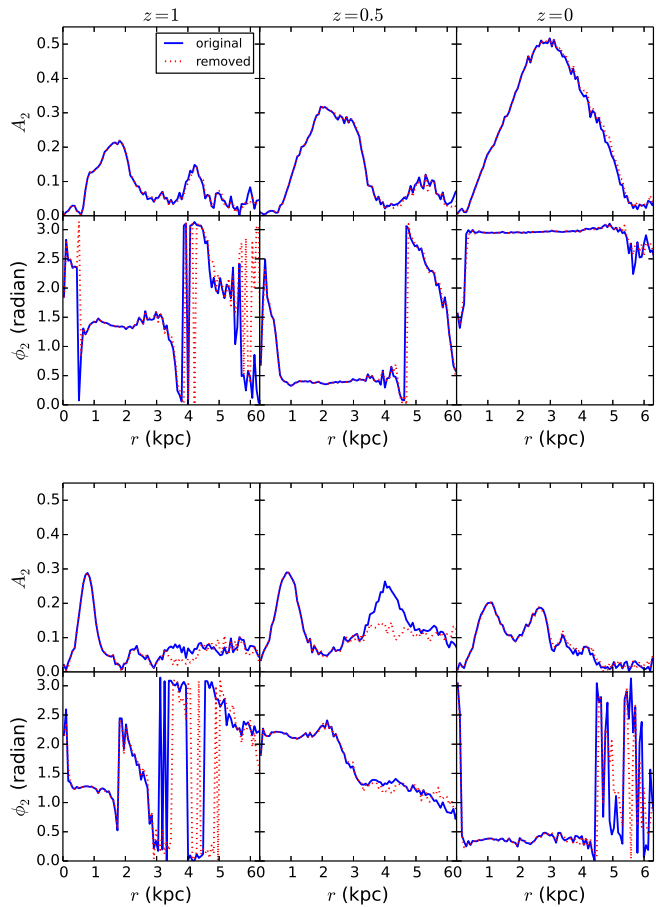} 
  \end{center}
  \caption{The same as figure~\ref{fig:m2} but here we compare the amplitude 
    and phase profiles from the original surface stellar density fields 
    (blue solid liens) with those from the density fields in which the 
    clumps have been removed (red dotted lines). 
}
\label{fig:m2comp}
\end{figure}

In figure~\ref{fig:m2comp}, we compare the amplitude and phase profiles 
of the $m = 2$ mode obtained from the original surface stellar density 
fields with those obtained from the density field in which 
the clumps have been removed. 
We find that the profiles are almost indistinguishable when 
there is no irregular structure like clumps. Hence the procedure described 
above does not affect our results when the clumps do not exist. 

At $z = 0.5$ in Aq-D, the clumps shown in figure~\ref{fig:contour_orig} 
contribute to the $m = 2$ mode. Removing the clumps lowers the second peak 
in the amplitude profile.
The phase profile is however hardly affected by the clump removal, 
since the arm-like features induced by the clumps remain present in 
the surface stellar density (see figure~\ref{fig:contour}). 

We have confirmed that the removal of the clumps does not 
affect our estimation of the bar length, amplitude, and phase 
for all the snapshots used in the paper. 
Nevertheless, we apply the clump removal procedure all the analyses 
present in this paper, unless otherwise stated. 

\section{Bar length} \label{app:bar_length}

\citet{am02} presented a number of ways to measure bar length. 
\citet{scannapieco12} tested three of them, which are applicable 
to cosmological simulations. 
We here use these three methods and see the uncertainties in 
measuring the bar length. 

The first method is the one we employ in the main text, which utilizes  
the radial phase profile of the $m = 2$ Fourier component.  
This method gives the bar length 5.58~kpc for Aq-C and 4.42~kpc 
for Aq-D at $z = 0$. 

In the second method, we utilize the amplitude of the $m = 2$ Fourier 
component. We estimate the bar length from the radius where the 
$m = 2$ amplitude drops to a small fraction of $A_2^\mathrm{max}$. 
We here define the bar length as the radius where $A_2(r)$ drops to
$0.25 A_2^\mathrm{max}$ as done in \citet{scannapieco12}. 
The bar lengths given by this method are 5.17~kpc and 4.42~kpc 
for Aq-C and Aq-D, respectively, at $z = 0$. 

\begin{figure}
  \begin{center}
  \includegraphics[width=\linewidth]{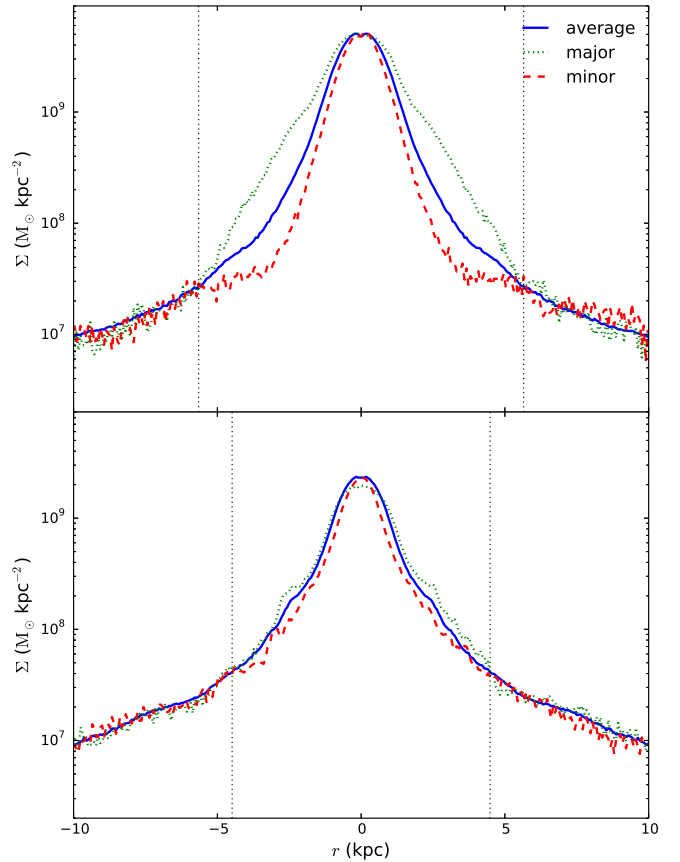} 
  \end{center}
  \caption{Surface stellar density profiles at $z = 0$, when 
    galaxies seen face-on. 
    The upper and lower panels respectively show Aq-C and Aq-D.   
    The azimuthally averaged profile is shown by the blue solid line, 
    and the profiles along the bar major axis and minor axis are 
    shown by the green dotted and red dashed lines, respectively. 
    The vertical dotted lines indicate the bar lengths defined by the 
    surface stellar density profiles. 
}
\label{fig:surfpro}
\end{figure}

The third method utilizes the surface stellar density profiles shown 
in figure~\ref{fig:surfpro}.
Within the bar length, the surface density along the bar major axis 
must be higher than that along the minor axis at given radius.  
For Aq-C, we can see the clear signature of the bar in the surface density 
profiles. 
On the other hand, the signature is much weaker for Aq-D. 
This is consistent with the small amplitude of the $m = 2$ Fourier mode in 
Aq-D. 
We define the bar length as the radius where the relative difference between 
the surface densities along the major and minor axes drops to 10 \% of
the maximum. 
This method yields the bar length of 5.65~kpc for Aq-C and 4.49~kpc for Aq-D 
as indicated in figure~\ref{fig:surfpro} by the vertical lines. 

The bar lengths obtained by the three different methods agree reasonably 
well between them both for Aq-C and Aq-D. 
We thus only show the bar length obtained by the first method in the 
main text.   

\section{The outer component of Aq-D's bar} \label{app:outer}
\begin{figure}
  \begin{center}
  \includegraphics[width=\linewidth]{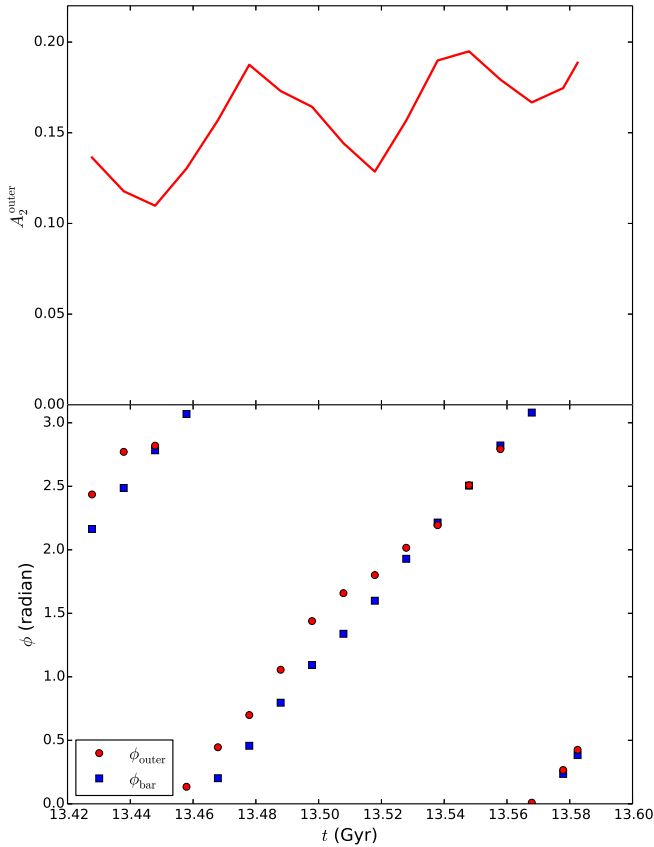} 
  \end{center}
  \caption{{\it Upper panel}: the amplitude of the $m = 2$ component 
    in Aq-D at the radius corresponding to the outer peak in the 
    amplitude profile. 
    {\it Lower panel}: The phase of the outer peak in the 
    amplitude profile of the $m = 2$ component of Aq-D (red filled points). 
    The phase of the bar, i.e. the phase at the radius corresponding to 
    the inner peak in the amplitude profile, is indicated by the blue filled 
    squares. 
}
\label{fig:outer}
\end{figure}

As we discussed in the main text, Aq-D's bar at $z = 0$ has two peaks 
in the amplitude of the $m = 2$ component within the bar length. 
This is quite unusual and thus we here present the evolution and 
properties of the outer component. 

In figure~\ref{fig:outer}, we show the evolution of the amplitude and 
the phase of the outer $m = 2$ component of Aq-D around $z = 0$. 
We find that the amplitude largely varies with time, while the amplitude 
of the inner peak does not change much (see figure~\ref{fig:pattern}). 

We also show the phase of the outer $m = 2$ component. 
Clearly, the outer component is almost always aligned with the inner 
component, though the slight misalignment at $t < 13.53$~Gyr is 
too large to include the outer component in the bar. 
Therefore, the outer component might be a part of the bar whose  
amplitude is enhanced and phase is shifted by the interactions 
with other components.

We show the face-on surface stellar density maps of Aq-D at 
$t \simeq 13.54$~Gyr when the amplitude of the outer $m = 2$ component 
is the largest. To see what is happening more closely, we draw contours 
that indicate the densities at radii corresponding to the minimum 
and outer peak of $A_2(r)$. 
The contours for the original density distribution clearly show the 
existence of substructure at the radius corresponding to the 
outer peak of $A_2(r)$. 
We always find such substructure when the amplitude of the outer component 
takes the maxima. 
The high frequency oscillation seen in the time evolution of the amplitude 
of the outer peak (see figure~\ref{fig:outer}) suggests that the outer peak 
is induced by objects that are orbiting in the disk with different angular 
velocity from the bar. 
We suspect that these rotating objects are debris of clumps from the surface
density contours. 

\begin{figure}
  \begin{center}
  \includegraphics[width=\linewidth]{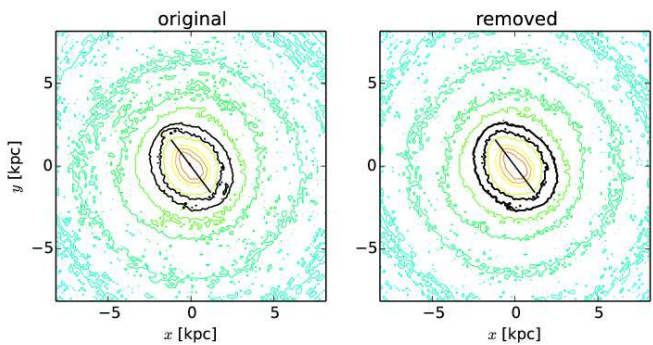} 
  \end{center}
  \caption{
    Contour maps of the face-on stellar surface density of Aq-D 
    when the amplitude of the outer $m = 2$ component is maximum 
    ($t \simeq 13.54$~Gyr). 
    The left and right panels respectively show the original  
    density field and the one in which the clumps have been removed. 
    The inner thick black solid contour corresponds to the densities 
    at the radius where $A_2(r)$ takes the minimum value between two peaks. 
    The outer thick black contour 
    indicates the density at the radius of the outer peak.  
    Here the bar length (the black straight line) is terminated at 
    the radius where $A_2(r)$ takes the minimum value between two peaks. 
}
\label{fig:contouter}
\end{figure}

After we have removed the clumps, we do not see such substructure in 
the density distribution (right panel of figure~\ref{fig:contour}). 
The amplitude of the $m = 2$ component however has a peak at the same 
radius even if we have removed the clumps because our procedure to 
remove the clumps do not erase the features induced by interactions 
with the clumps. 
When the amplitude of the outer component is smallest, the amplitude profile 
of the $m = 2$ component is similar to that of a weak bar whose length is 
beyond the radius of the outer peak, although the amplitude profile is very 
noisy.    
We therefore do not exclude the outer $m = 2$ component from the bar 
as long as its phase is aligned with $\phi_\mathrm{bar}$, i.e. 
the phase of the inner peak. 
We stress that the bar angle, $\phi_\mathrm{bar}$, and the bar 
amplitude, $A_2^\mathrm{max}$, are not affected by the inclusion 
(or exclusion) of the outer component because we always employ 
the radius of the inner peak as $r_2^\mathrm{max}$.

%

\end{document}